\documentclass[twocolumn, pra, a4paper, 10pt, superscriptaddress, nopacs, floatfix]{revtex4-1}
\usepackage{siunitx}
\usepackage{graphicx}
\usepackage{xcolor}
\usepackage{physics}
\usepackage{nicefrac}
\usepackage{float}
\usepackage[utf8]{inputenc}

\begin{document}

\title{Single photon wavefront-splitting interference : \\
an illustration of the light quantum in action}

\author{V. Jacques}
\affiliation{Laboratoire de Photonique Quantique et Mol\'eculaire ENS Cachan, UMR CNRS 8537, Cachan, France}
\author{E Wu}
\affiliation{Laboratoire de Photonique Quantique et Mol\'eculaire ENS Cachan, UMR CNRS 8537, Cachan, France}
\affiliation{Key Laboratory of Optical and Magnetic Resonance Spectroscopy, East China Normal University, Shanghai, China}
\author{T. Toury}
\affiliation{Laboratoire de Photonique Quantique et Mol\'eculaire ENS Cachan, UMR CNRS 8537, Cachan, France}
\affiliation{Palais de la Découverte, Paris, France}
\author{F. Treussart}
\affiliation{Laboratoire de Photonique Quantique et Mol\'eculaire ENS Cachan, UMR CNRS 8537, Cachan, France}
\author{A. Aspect}
\affiliation{Laboratoire Charles Fabry de l'Institut d'Optique, UMR CNRS 8501, Orsay, France}
\author{P. Grangier}
\affiliation{Laboratoire Charles Fabry de l'Institut d'Optique, UMR CNRS 8501, Orsay, France}
\author{J.-F. Roch}
\affiliation{Laboratoire de Photonique Quantique et Mol\'eculaire ENS Cachan, UMR CNRS 8537, Cachan, France}

\begin{abstract}
We present a new realization of the textbook experiment consisting in single-photon interference based on the pulsed, optically excited photoluminescence of a single colour centre in a diamond nanocrystal. Interferences are created by wavefront-splitting with a Fresnel's biprism and observed by registering the "single-photon clicks" with an intensified CCD camera. This imaging detector provides also a real-time movie of the build-up of the single-photon fringes. We perform a second experiment with two detectors sensitive to photons that follow either one or the other interference path. Evidence for single photon behaviour is then obtained from the absence of time coincidence between detections in these two paths.  
\end{abstract}

\maketitle
\section{Introduction}
\label{intro}
Many textbooks on Quantum Mechanics, like Feynman's lectures \cite{Feynman}, describe a thought-experiment based on the Young's double-slit interference setup, and realised with independent particles sent one at a time through the interferometer. The striking feature is that the image we have for interference is a wave passing simultaneously through both slits, incompatible with the image of a particle, which goes through either one or the other slit but not both. While the last statement comes naturally for objects primarily known as particles like electrons \cite{Tonomura}, neutrons \cite{Neutrons}, atoms \cite{Carnal} and molecules  \cite{Arndt}, it can be questioned in the case of the ''LichtQuanten" introduced by Einstein \cite{Einstein} in 1905, since light is primarily described as a wave. 

\begin{figure}
\includegraphics[width=8.5cm]{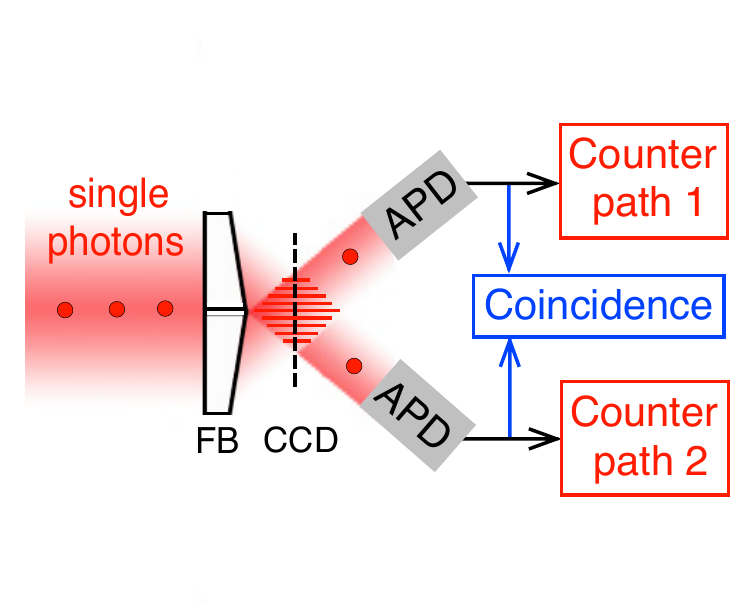}
 \caption{Wavefront-splitting setup based on a Fresnel's biprism (FB). APDs are avalanche silicon photodiodes operating in photon counting regime. An intensified CCD camera (dash line) records interference fringes in the overlapping region of the two deviated wavefronts. When the CCD is removed, it is then possible to demonstrate the single photon behaviour by recording the time coincidences events between the two output channels of the interferometer.}
 \label{general}
 \end{figure}
 
One century later, we present a realization of this textbook experiment consisting in single-photon interference. Our experiment, depicted on Fig.\ref{general}, has several new striking features compared to previous works \cite{Grangier,Zeilinger,Benson,Jelezko}: (i) we use a clock-triggered single-photon source from a single emitting dipole, which is both conceptually and practically simple \cite{Kun,Brouri}, (ii) we use a wavefront-splitting interferometer based on a Fresnel's biprism, very close to the basic Young's double-slit scheme, (iii) we register the ''single-photon clicks'' in the interference plane using an intensified CCD camera which provides a real-time movie of the build-up of the single-photon fringes. 

In early experiments \cite{Taylor}, the so-called single-photon regime was reached by attenuating light. But Poissonian photon number statistics in faint light pulses leads to the unwanted feature that more than one photon may be present between the source and the interference fringes observation plane. To demonstrate the single-photon behaviour, we also perform the experiment with two detectors sensitive to photons that follow either one or the other interference path. Evidence for single photon behaviour can then be obtained from the absence of time coincidence between detections in these two paths \cite{Grangier,Gerry}. Such a measurement also provides a ''which-path'' information complementary to the interference observation.

The paper is organized as follows. In section $2$, we describe the triggered single-photon source used for the experiment. Section 3 is dedicated to the demonstration of the particle-like behaviour. Finally, single photon interferences are presented in section 4. 

\section{Triggered single-photon source using the photoluminescence of a single N-V
colour centre in a diamond nanocrystal }
\label{sec:1}

A lot of efforts have been put in the realization of single-photon-source (SPS) over the recent years. Since first proposal \cite{Imamoglu}, a wide variety of schemes have been worked out, based on the fluorescence from different kinds of emitters, such as molecules, atoms, colour centres and semiconductor structures \cite{General SPS}.
The clock-triggered single-photon source at the heart of our experiment, previously developed for quantum key distribution \cite{Brouri,Alleaume}, is based on the pulsed, optically excited photoluminescence of a single N-V colour centre in a diamond nanocrystal. This system, which consists in a substitionnal nitrogen atom (N) associated to a vacancy (V) in an adjacent lattice site of the diamond crystalline matrix (Fig.2-(a)), has shown an unsurpassed efficiency and photostability at room temperature \cite{Gruber,Kurtsiefer}. 

\begin{figure}[h!]
  \centerline{\includegraphics[width=8.8cm]{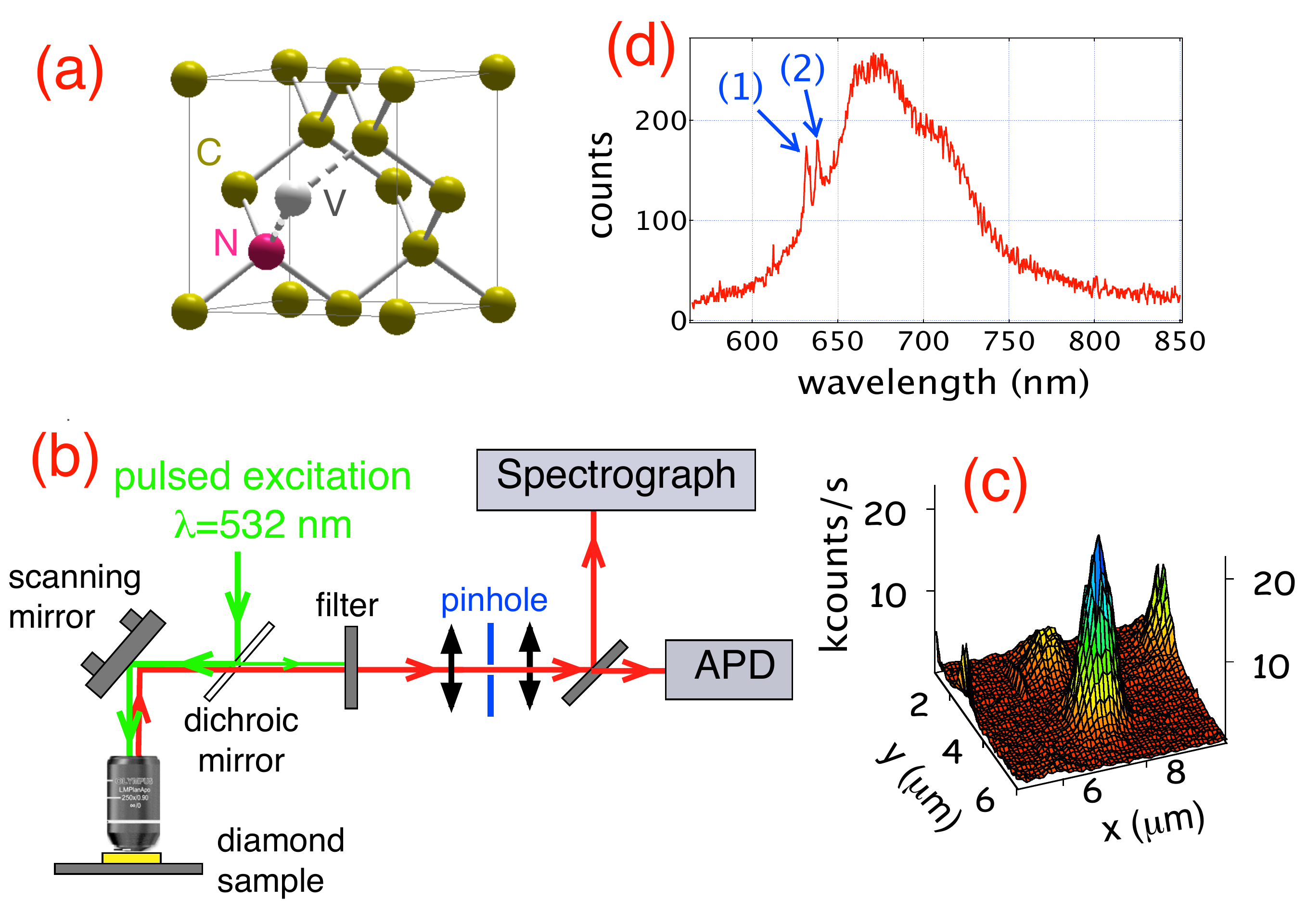}}
  \caption{(a)-The N-V  colour centre  consists in a substitionnal nitrogen atom (N), associated to a
 vacancy (V)  in an adjacent lattice site of the diamond crystalline matrix. 
(b)-Confocal microscopy setup: pulsed excitation laser
beam at $\lambda =  532$ nm is tightly focused on the
diamond nanocrystals with a high numerical aperture (NA=$0.95$) microscope
objective. Fluorescence light emitted by the N-V colour centre
is collected by the same objective and then spectrally filtered from the
remaining pump light. Following  standard
confocal detection scheme, collected light is focused onto a $100$
microns diameter pinhole. 
The reflected part from a $50$/$50$ beamsplitter (BS) goes then into an imaging spectrograph,
while the transmitted part is detected by a silicon avalanche photodiode in photon counting regime and used for fluorescence raster scan as shown in (c). The central peak corresponds to the 
photoluminescence of the single N-V colour centre used in the single-photon interference experiment. (d)-Fluorescence spectrum of this single N-V colour centre recorded with a  back-illuminated cooled CCD matrix (2 minutes integration duration)  placed in the image plane of the spectrograph. The peak fluorescence wavelength is at
670~nm. The two sharp lines (1) and (2) are respectively the two-phonon Raman
scattering line of the diamond matrix associated to the excitation wavelength
and the N-V centre zero phonon line at 637~nm which characterizes the negatively charged N-V colour centre photoluminescence.}
\label{confoc}
\end{figure}

In bulk diamond, the high index of refraction of the material
($n=2.4$) makes difficult to extract N-V colour centre fluorescence efficiently. One way to circumvent this problem is to use diamond nanocrystals, with a size much smaller than the radiated light wavelength, deposited on a microscope glass coverplate~\cite{Beveratos}.  For such sub-wavelength size, refraction becomes irrelevant and the colour centre can
simply be assimilated to a point source radiating at the air-glass
interface. Samples were prepared by a procedure described in Ref.\cite{Beveratos}. N-V colour centre are created by irradiation of type Ib diamond powder with high-energy electrons followed by annealing at $800^{\circ}$C. Under well controlled irradiation dose, the N-V colour centre density
is small enough to allow independent
addressing of a single emitter using standard confocal microscopy, as depicted on Fig.\ref{confoc}-(b).

Under pulsed excitation with a pulse duration shorter than the
radiative lifetime, a single dipole emits photons one by one \cite{Brouri_PRA00,De Martini}. As described in Ref.\cite{Kun}, we use a home-built pulsed laser at a wavelength of 532~nm with a 800~ps pulse duration to excite a single N-V colour centre. The 50~pJ energy per pulse is high enough to ensure efficient pumping of the defect centre in its excited level. The
repetition rate, synchronized on a stable external clock, can be adjusted
between 2 to 6 MHz so that successive fluorescent decays are well
separated in time from each other. Single photons are thus emitted by
the N-V colour centre at predetermined times within the accuracy of
its excited state lifetime, which is about 45~ns (See Fig.\ref{fig_biprisme_HBT}) for the centre used in
the experiment.

As a proof of the robustness of this single-photon source, note that all the following experiments have been realised with the same emitting N-V colour centre.

 \section{"Which-path" experiment : particle-like behaviour}
Single photons emitted by the N-V  colour centre are now sent at normal incidence onto a
Fresnel's biprism. Evidence for a particle-like behaviour can be obtained using the arrangement of Fig.\ref{general}. If light is really made of quanta, a single photon should either be deviated upwards or downwards, but should not be split by the biprism. In that case, no coincidences corresponding to joint photodetections on the two output beams should be observed. On the opposite, for a semi-classical model that describes light as a classical wave, the input wavefront will be split in two equal parts, leading to a non-zero probability of joint detection on the two photodetectors. Observation of zero coincidences, corresponding to an anticorrelation effect, would thus give evidence for a particle-like behaviour. 
 
For a realistic experiment aimed at evidencing this property, we need to establish a criterion which enables us to discriminate between a particle-like behaviour and another one compatible with the semi-classical model for light. For that, we faithfully follow the approach introduced in Ref.\cite{Grangier} for interpreting single photon anticorrelation effect on the two output channels of a beamsplitter.  We consider a run corresponding to $N_{T}$ trigger pulses applied to the emitter, with $N_{1}$ (resp. $N_{2}$) counts detected in path 1 (resp. 2) of the interferometer, and $N_{C}$ detected coincidences. It is straightforward to show that any semi-classical theory of light, in which light is treated as a wave and photodetectors are quantized, predicts that these numbers of counts should obey the inequality 

\begin{equation} 
\alpha = \frac{N_{C}N_{T}}{N_{1}N_{2}} \geq 1.
\end{equation} 

Violation of this inequality thus gives a criterion which characterizes nonclassical behaviour. For a single photon wavepacket, perfect anticorrelation is predicted since the photon can only be detected once, leading to $\alpha=0$ in agreement with the intuitive image that a single photon cannot be detected simultaneously in the two paths of the interferometer. On the other hand, inequality (1) cannot be violated even with faint laser pulses. Indeed, in this case the number of photons in the pulse follows a Poisson law predicting $\alpha=1$. This value indicates that coincidences will then be observed preventing the particle-like behaviour to be evidenced. We measured the $\alpha$ correlation parameter for triggered single-photon pulses and for faint laser pulses. To establish a valid comparison between these two cases, all data have been taken for an identical mean number of detected photons per pulse, below $10^{-2}$. 

\begin{figure}[t]
 \centerline{\includegraphics[width=9cm]{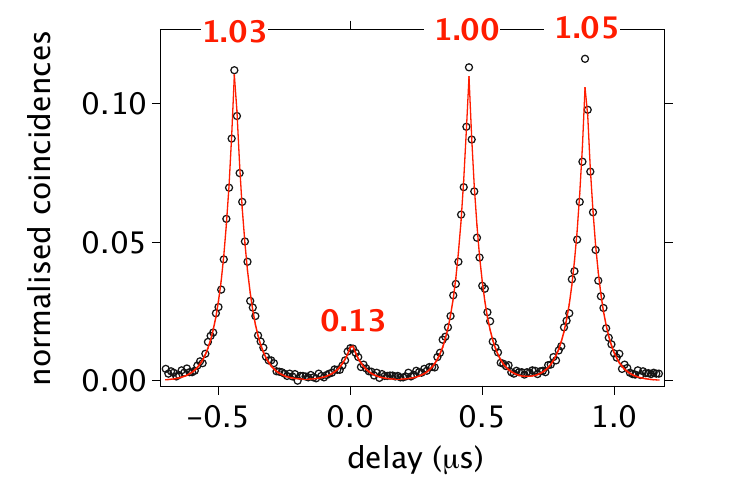}}
 \caption{Histogram of time intervals between consecutive photodetection events in the two paths of the wavefront-splitting interferometer.  The pulsed excitation laser repetition period is 444~ns, and the total counting time was 148.15~s . Lines are exponential fits for each peak, corresponding to a radiative lifetime of  $44.6\pm 0.8$~ns. Values written above each peak correspond to their respective  area normalized to the corresponding value for poissonian photon number statistics. The strong reduction of coincidences around zero delay gives evidence for single-photon emission by the excited N-V colour centre and for   its particle-like behaviour as the wavefront  is split in two parts by the Fresnel's  biprism.}
 \label{fig_biprisme_HBT}
\end{figure}

\begin{table}
\caption{Measurements of the correlation parameter $\alpha$ associated to ten independant sets of $10^5$ photodetections (lasting about 5~s) registered for ({\it a})  faint laser pulses with a mean number of photons per pulse below $10^{-2}$ and ({\it b}) single-photon pulses emitted by the N-V colour centre.
As the laser emits coherent states of light, the number of photons in each pulse is given by Poissonian statistics, leading to a correlation parameter $\alpha = 1.00 \pm 0.06$, the precision being inferred from simple statistical analysis with a 95\% confidence interval. On the other hand, an anticorrelation effect, corresponding to $\alpha = 0.13 \pm 0.01<1$, is clearly observed for single-photon pulses propagating through the Fresnel's biprism.}
\label{tab:1}      

\begin{center}
({\it a}) {\it Faint laser pulses}\\
\vspace{0.15cm}
\begin{tabular}{lllll}
\hline\noalign{\smallskip}
Counting time (s) 
    & $N_{1}$ 
      &  $N_{2}$
          &   $N_{C}$ 
             &   $\alpha$\\
             \noalign{\smallskip}\hline\noalign{\smallskip}
4.780 & 49448 & 50552 & 269 & 1.180\\
          4.891 & 49451 & 50449 & 212 & 0.937\\
        4.823 & 49204 & 50796 & 211 & 0.934\\
          4.869 & 49489 & 50511 & 196 & 0.875\\
          4.799 & 49377 & 50623 & 223 & 0.981\\
          4.846 & 49211 & 50789 & 221 & 0.982\\
          4.797 & 49042 & 50958 & 232 & 1.021\\
          4.735 & 49492 & 50508 & 248 & 1.077\\
          4.790 & 49505 & 50495 & 248 & 1.090\\
          4.826 & 49229 & 50771 & 219 & 0.970\\

\noalign{\smallskip}\hline
\end{tabular}
\end{center}

\vspace{0.35cm}
\begin{center}
({\it b}) {\it Single-photon pulses}\\
\vspace{0.15cm}
\begin{tabular}{lllll}
\hline\noalign{\smallskip}
Counting time (s) 
    & $N_{1}$ 
      &  $N_{2}$
          &   $N_{C}$ 
             &   $\alpha$\\
             \noalign{\smallskip}\hline\noalign{\smallskip}
       
         5.138 & 49135 & 50865 & 28 &  0.132\\
          5.190 & 49041 & 50959 & 23 & 0.109\\
        5.166 & 49097 & 50903 & 23 & 0.109\\
          5.173 & 49007 & 50996 & 28 & 0.133\\
          5.166 & 48783 & 51217 & 29 & 0.137\\
          5.167 & 48951 & 51049 & 31 & 0.147\\
          5.169 & 49156 & 50844 & 30 & 0.142\\
          5.204 & 49149 & 50851 & 32 & 0.152\\
          5.179 & 49023 & 50977 & 26 & 0.124\\
          5.170 & 48783 & 51217 & 26 & 0.123\\
\hline
\end{tabular}
\end{center}
\end{table}

Since the single-photon source emits light pulses triggered by a stable external clock and
well separated in time,  the value of $\alpha$ can be directly inferred from the record of all photon arrival times. Every photodetection event produced by the two avalanche photodiodes  is time-stamped using a time-interval-analyser (TIA) computer board 
(GT653, GuideTech).  Straightforward processing of these timestamps over a discrete time base allows us to reconstruct the number of detection events on each output channel of the biprism interferometer, and thus gives an access to the "which-path" information. Furthermore,  time intervals between two successive photodetections are directly inferred from the timestamps ensemble,  so as to estimate if  a coincidence has occurred or not for each registered photodetection. However  the notion of coincidence is meaningful only accordingly to a temporal gate: there will be a coincidence if two detections happen within the same gate. As the radiative lifetime of the emitting N-V colour centre 
used for the experiment is approximately equal to 45~ns (see Fig.\ref{fig_biprisme_HBT}),   we set the   gate duration to 100~ns. Such a value is  much smaller than the 436~ns time interval between two successive excitation pulses. It  also ensures that about  90\% of the detected photons are considered for data analysis. 

Using the results given on Table~1
 and simple statistical analysis associated to a 95\% confidence interval, we infer 
$\alpha = 1.00 \pm 0.06$  for faint laser pulses and 
$\alpha = 0.13 \pm 0.01$ for single-photon pulses. 

As expected, light pulses emitted by the single-photon source lead  to a  strong violation of the inequality $\alpha \geq 1$ valid for any semi-classical theory of light. The non-ideal value  $\alpha \not= 0 $ is due to remaining background fluorescence of the sample and to the two-phonon Raman scattering line, which induces a non-vanishing probability of having more than one photon in the emitted light pulse.  Table~1 also shows that the number of coincidences observed with faint laser pulses, within a given integration time, is much higher than in the experiment with single-photon pulses, and corresponds to $\alpha$ equal to unity within one standard deviation. This result  is a clear confirmation that it is not possible to demonstrate the particle-like behaviour with attenuated pulses from a classical light source, involving many emitters simultaneously and independently excited.    

From the set of photodetection timestamps, one can also build the histogram of delays between two consecutive detections on the two paths of the wavefront-splitting  interferometer.  In the limit of low collection efficiency and short timescale, the recorded histogram coincides with a measurement of the second-order  autocorrelation function $g^{(2)}$ \cite{Reynaud}. 
As shown in Fig.\ref{fig_biprisme_HBT}, strong reduction of coincidences around zero delay gives again clear evidence for single-photon behaviour \cite{Brunel}. Note that the area of the zero-delay peak, normalized to its value for coherent pulses with poissonian statistics, is strictly equivalent to the $\alpha$ correlation parameter previously defined.

\section{Single-photon interference setup and data analysis}
\label{sec:3}

\subsection{Experimental setup and observations}
\label{sec:2.1}
 Using the same N-V colour centre, we now observe the interference fringes in the
intersection volume of the two separated wavefronts, as shown in Fig.\ref{general}. 

Interference patterns are detected in a single-photon recording mode,
using an intensified CCD camera (\emph{iStar} from
Andor Technologies, Ireland) cooled at $-25^\circ{\rm C}$. An eye-piece, equivalent
to a short focal length achromatic lens ($f=22$~mm)  is inserted between the biprism and the camera, 
in order to obtain a fringe spacing much larger than the $25~\mu$m  pixel size of the camera. 
Image acquisition parameter are optimized by adjusting the camera gain and the detection threshold, so that more than $\approx 88\%$ of the bright pixels correspond to detections of single photons.
 
 \begin{figure}[ht]
  \centerline{\includegraphics[width=9cm]{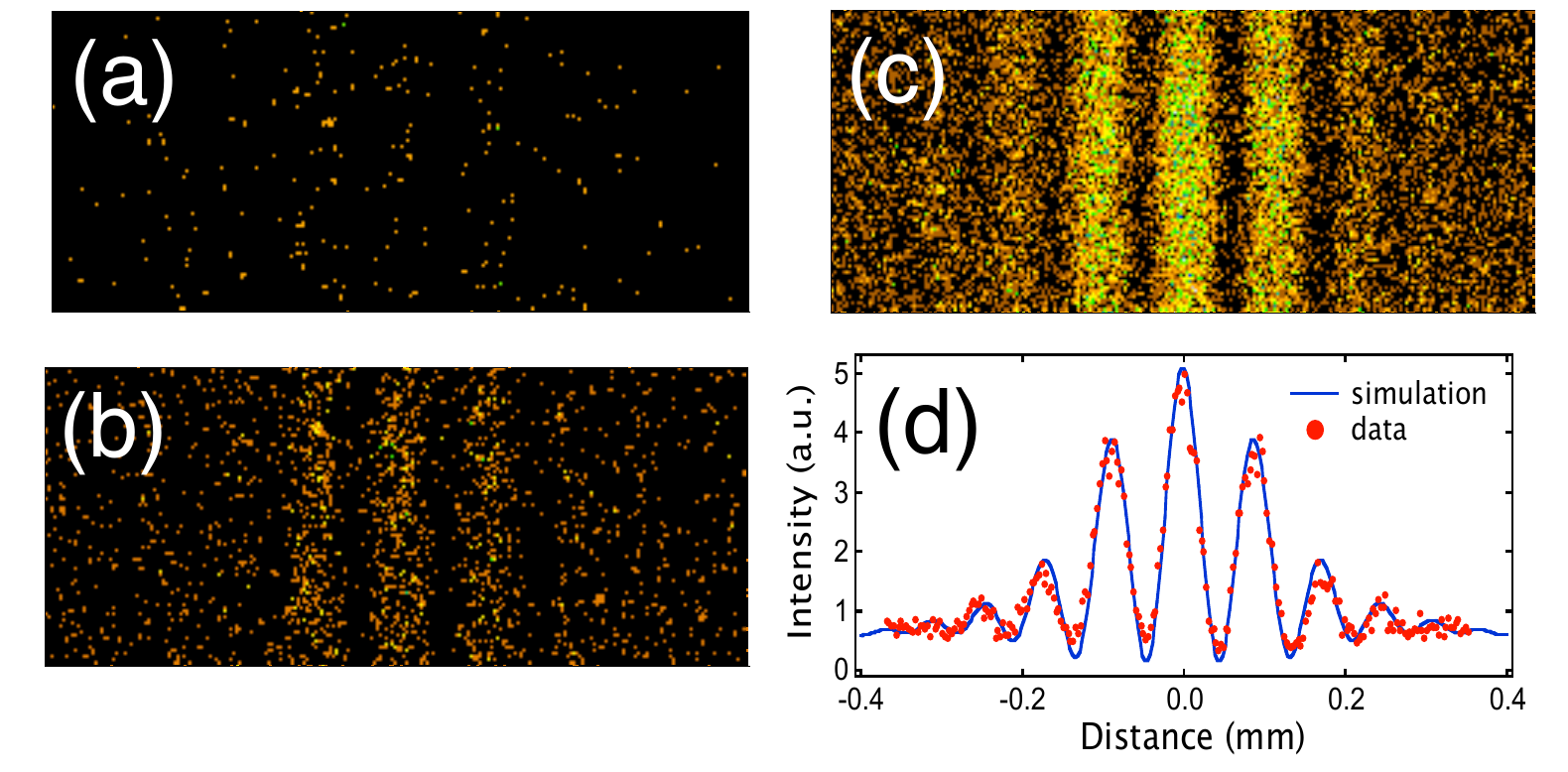}}
 \caption{Observation of the interference pattern expanded by the eyepiece and recorded by the intensified CCD camera. Image (a) (resp. (b) and (c)) is made of 272 photocounts (resp. 2240 and 19773)  corresponding to an exposure duration of 20~s (resp. 200~s  and 2000~s). Graph (d) displays the resulting interference fringes obtained by binning columns of CCD image (c) and fit of this interference pattern using coherent beam propagation in the Fresnel diffraction regime, and taking into account the finite temporal coherence due to the broad spectral emission of the NV colour centre. A visibility of 94\% can be associated to the central fringe.}
\label{interf_setup}
\end{figure}

Snapshots with exposure duration of 1~s are acquired one after each
other, with a mean number of eight photons detected on the CCD
array per snapshot. Accumulated images show how
the interference pattern builds up (Fig.\ref{interf_setup}). Less than 200 accumulated snapshots are required to clearly see the fringes pattern, and the final image contains about $2\times 10^4$ single photodetection events. This is a demonstration of the wave-like behaviour of light, even in the single-photon regime. To display fringe pattern gradual build-up, we generated a movie from the 2000 snapshots (available as Supplementary File). The movie frame rate is set to a value of  30 images per second resulting in a build-up 30 times faster than in real time.

\subsection{Fit of the interference pattern}
\label{sec:2.2}
The interference pattern obtained with a monochromatic plane wave incident on Fresnel's biprism would result in sinusoidal fringes with a visibility equal to unity. In our experiment,  the finite spatial extension of the single-photon wavefront yields more complicated interference patterns. In order to fit the fringes shown on Fig.\ref{interf_setup}-(c), we performed a computer simulation of the beam propagation through the Fresnel's biprism using free space propagation theory from classical optics. 
Starting at the input of the Fresnel's biprism with a TEM$_{00}$ wavefront associated to a measured 1.25~mm FWHM gaussian intensity distribution, the output field amplitude is calculated in a virtual observation plane located at a given distance $z$ behind the biprism input plane~\cite{Note}. The expected pattern in the CCD plane simply results
from the expansion by the eye-piece of this calculated intensity pattern. 

In order to fit more accurately the fringes minima, we also take into account temporal coherence effects due to the N-V colour centre broadband incoherent emission (see Fig.\ref{confoc}-(d)).  Finally the calculated intensity pattern is normalized to the total recorded intensity so that the only remaining fitting parameter is the observation distance $z$. 
The quality of the fit is illustrated by Fig.\ref{interf_setup}-(d), recorded approximately in the middle of the interference field. All over the interference pattern, the visibility of the fringes is very well described by the beam propagation simulation with the addition of finite temporal coherence of the source.

  \begin{figure} [t]
  \centerline{\includegraphics[width=8.8cm]{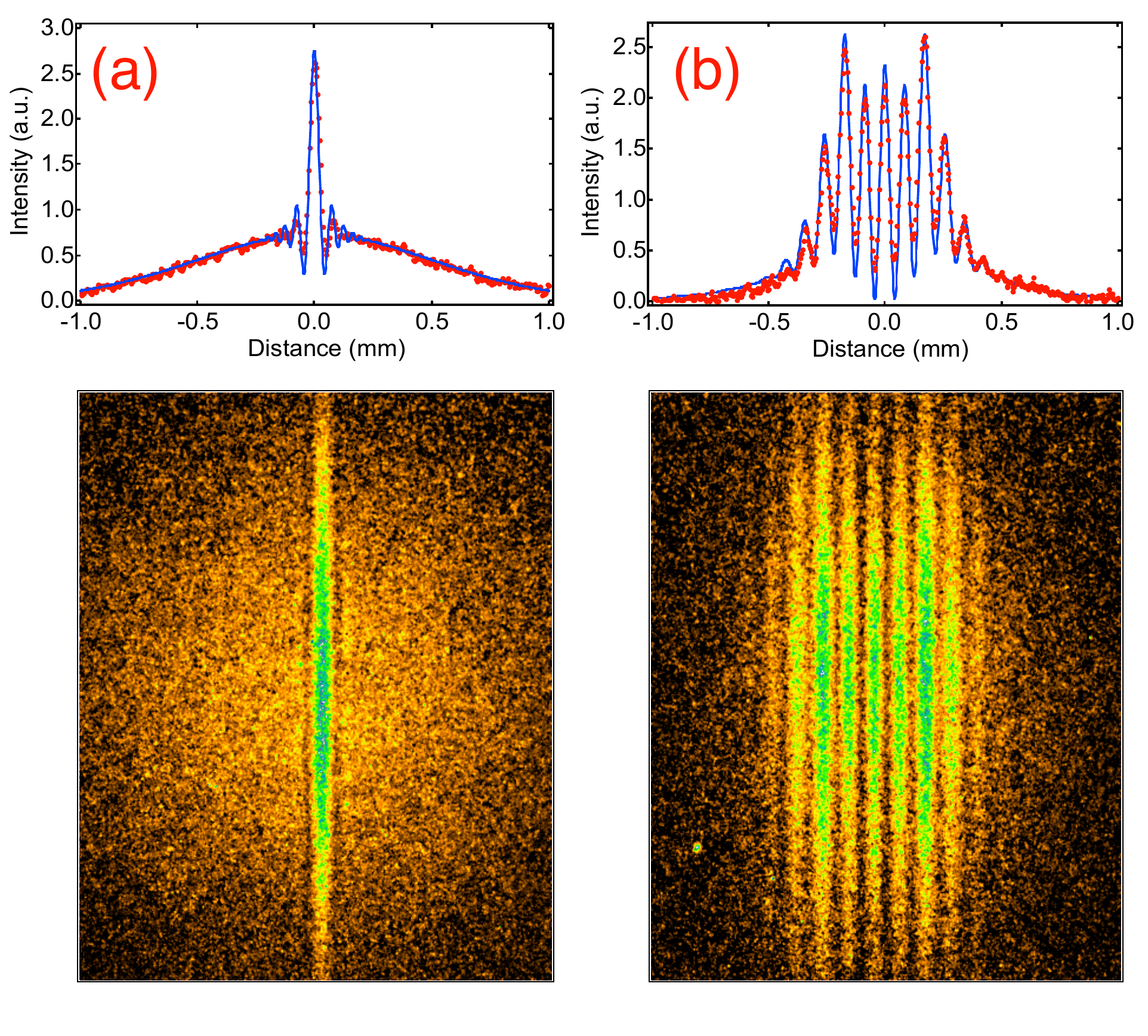}}
  \caption{Examples of interference patterns observed in the overlapping region of the two transmitted beams through the Fresnel's biprism, for (a) $z = 11 $ mm and (b) $z = 98$ mm.  Overlayed blue curves are fits evaluated from a model taking   into account Fresnel diffraction and temporal coherence effects.}
 \label{fresnel}
 \end{figure}

Validity of the model is furthermore confirmed by fits of interference patterns
observed for two others $z$ positions, as shown in Fig.\ref{fresnel}.
The good agreement between the fit and the experimental data is an illustration of the well-known result from Quantum Optics that phenomena like interference, diffraction, propagation, can be computed with the classical theory of light even in the single-photon regime \cite{Gerry}. 

\section{Conclusion}
  \label{sec:conc}

Summarizing, we have carried out the two complementary experiments of "interference observation vs. which-path detection" in the single-photon regime. While our results may not appear as a big surprise, it is interesting to note that one century after Einstein's paper the intriguing properties of the photon still give rise to sometimes confused debates \cite{Afshar}. We hope that our experiment can contribute not only to clarify such discussions, but also to arouse the interest and astonishment of those who will discover the photon during the century to come.
%

\begin{thebibliography}{}

\bibitem{Feynman}
  R. P. Feynman, R. B. Leighton,  Sands M. L., \textit{Lectures on Physics}, (Addison Wesley, Reading, 1963)
   
 \bibitem{Tonomura}
A. Tonomura, J. Endo, T. Matsuda, T. Kawasaki, and H. Ezawa, Am. J. Phys. \textbf{57}, 117 (1989)

\bibitem{Neutrons}
J. Simmhammer, G. Badurek, H. Rauch, U. Kischko, and A. Zeilinger, Phys. Rev. A \textbf{27}, 2523 (1983)
 
 \bibitem{Carnal}
O. Carnal and J. Mlynek, Phys. Rev. Lett. \textbf{66}, 2689 (1991)
 
 \bibitem{Arndt}
M. Arndt, O. Nairz, J. Vos-Andreae, C. Keller, G. van der Zouw, and A. Zeilinger, Nature \textbf{401}, 680 (1999)
 
   \bibitem{Einstein}
A. Einstein, Ann. d. Phys. \textbf{17}, 132 (1905)

 \bibitem{Grangier}
P. Grangier, G. Roger and A. Aspect,  Europhys. Lett. \textbf{1}, 173 (1986)

\bibitem{Zeilinger}
A. Zeilinger, G. Weihs, T. Jennewein, and M. Aspelmeyer, Nature \textbf{433}, 237 (2005)

\bibitem{Benson}
T. Aichele, U. Herzog, M. Scholtz, and O. Benson, e-print quant-ph/0410112

\bibitem{Jelezko}
F. Jelezko, A. Volkmer, I. Popa, K. K. Rebane, and J. Wrachtrup, Phys. Rev. A \textbf{67}, 041802 (2003)

\bibitem{Kun}  
A. Beveratos, S. Kuhn, R. Brouri, T. Gacoin, J.-P. Poizat, and P. Grangier, Eur. Phys. J. D \textbf{18}, 191 (2002)

\bibitem{Brouri}
A. Beveratos, R. Brouri, T. Gacoin, A. Villing , J.-P. Poizat, and P. Grangier, Phys. Rev. Lett. \textbf{82}, 187901 (2002)

\bibitem{Taylor}
G. I. Taylor, Proc. Com. Philos. Soc. \textbf{15}, 114 (1909)


\bibitem{Gerry}
   C. Gerry and P. Knight, \textit{Introductory Quantum Optics}, (Cambridge University Press, 2004)
   
   
    \bibitem{Imamoglu}
    A. Imamoglu and Y. Yamamoto, Phys. Rev. Lett. \textbf{72}, 210 (1994)

   
\bibitem{General SPS}
See e.g. Focus on Single Photons on Demand, New J. Phys. \textbf{6} (2004)

\bibitem{Alleaume}
R. Alléume, F. Treussart, G. Messin, Y. Dumeige, J.-F. Roch, A. Beveratos, R. Brouri-Tualle, J.-P. Poizat, and P. Grangier, New J. Phys. \textbf{6}, 92 (2004)


\bibitem{Gruber}
A. Gruber, A. Drabenstedt, C. Tietz, L. Fleury, J. Wrachtrup, and C. von Borczyskowski, Science \textbf{276}, 2012 (1997)

\bibitem{Kurtsiefer}   
C. Kurtsiefer, S. Mayer, P. Zarda, and H. Weinfurter, Phys. Rev. Lett. \textbf{85}, 290 (2000).

\bibitem{Beveratos}
A. Beveratos, R. Brouri, T.
Gacoin, J.-P. Poizat, and P.
Grangier, Phys. Rev. A \textbf{64}, 061802R (2001)

\bibitem{Brouri_PRA00}
R.~Brouri, A.~Beveratos, J.-P.~Poizat and P.~Grangier,
Phys. Rev. A \textbf{62}, 063817 (2000)

\bibitem{De Martini}   
F. De Martini, G. Di Guiseppe, and M. Marrocco, Phys. Rev. Lett. \textbf{76}, 900 (1996)

\bibitem{Reynaud}
S. Reynaud, Ann. Phys. Fr \textbf{8}, 315 (1983)

\bibitem{Brunel}
C. Brunel, B. Lounis, Ph. Tamarat, and M. Orrit, Phys. Rev. Lett. \textbf{81}, 2679 (1998)

\bibitem{Note}
Numerical calculations were done using the \emph{LightPipe} routines for Mathcad written by Fred van Goor (Department of Applied Physics, University of Twente, The
Netherlands)

\bibitem{Afshar}
See e.g.   R.E. Kastner, The Afshar Two-Slit Experiment and Complementarity, e-print quant-ph/0502021

\end{thebibliography}
%

\end{document}